%% file: main.tex
\documentclass[11pt]{article}
\usepackage[T1]{fontenc}
\usepackage[utf8]{inputenc}
\usepackage[margin=1in]{geometry}
\usepackage{times,microtype,graphicx,array,amsmath,booktabs,longtable}
\usepackage{float}
\usepackage{pgfplots}
\pgfplotsset{compat=1.18}
\usetikzlibrary{arrows.meta,calc,positioning}
\usepackage{amssymb}
\definecolor{vermillion}{RGB}{213,94,0}
\definecolor{oiblue}{RGB}{0,114,178}
\definecolor{bluishgreen}{RGB}{0,158,115}
\usepackage[round,authoryear]{natbib}
\usepackage{xurl}
\usepackage[hidelinks]{hyperref}
\usepackage{authblk}

\newcommand{\framework}{\textsc{dsh}}
\input{generated/benchmark-metadata.tex}

\title{Evaluating Agent Skills for Version-Specific Plugin Migration: A Retrospective Study}
\author[1]{Beiming~Liu\thanks{Corresponding author: \texttt{lbm21@tsinghua.org.cn}}}
\author[2]{Haihao~Li\thanks{Equal contribution.}}
\author[3]{Minjie~Chen\protect\footnotemark[2]}
\author[4]{Ning~Chen}
\author[5]{Yiran~Wang}
\author[6]{Jiming~Ye}
\author[7]{Puzhao~Zhang}
\author[8]{Tongtao~Wang}
\author[9]{Sheng~Gao}
\author[8]{William~Jin}
\author[10,11]{Weihao~Mu}
\author[12]{Chengzhi~Liu}
\author[13]{Yucheng~Xia}
\author[14]{Guangren~Wang}
\author[15,16]{Chaoyang~Fan}
\author[17]{Changfeng~Huang}
\author[18]{Xunming~Lin}
\author[19]{Yuanjie~Shen}
\affil[1]{Tsinghua University}
\affil[2]{Fudan University}
\affil[3]{PetroChina Southwest Oil \& Gasfield Company}
\affil[4]{Jilin University}
\affil[5]{The Frederick Gunn School}
\affil[6]{Shenzhen University}
\affil[7]{Dalian Neusoft University of Information}
\affil[8]{Independent Researcher}
\affil[9]{University of Chinese Academy of Sciences}
\affil[10]{Dalian University of Technology}
\affil[11]{Jiyin Zhiyuan (Shanghai) Technology Co., Ltd.}
\affil[12]{Lanzhou University of Technology}
\affil[13]{Harbin Engineering University}
\affil[14]{Alibaba Cloud}
\affil[15]{Jianghan University}
\affil[16]{Jiuxiangxian (Beijing) Technology Co., Ltd.}
\affil[17]{Sun Yat-sen University}
\affil[18]{Great Bay University}
\affil[19]{Beijing Information Science and Technology University}
\date{}
\begin{document}
\maketitle
\begin{abstract}
Agent skills package version-specific maintenance knowledge for coding agents, but a higher diagnostic score does not by itself show that the resulting migration advice satisfies the target version's contract. We study a shipped plugin-upgrade skill through an archive of 64 reports on 16 static migration tasks, with two attempts per condition and 328 criterion decisions. With the skill, mean recorded reward rises from 93.83 to 98.75, a gain of 4.92 points (95\% task-bootstrap interval [0.31, 10.86]); the gain is concentrated in one task, and eight task pairs are at the ceiling. Tracing every decision to its contract domain and reviewing ten reports in depth exposes grading errors that favor either arm; in one, a containment predicate that accepts the parent directory still receives full credit. Executable probes confirm this defect and show that a working teardown repair is excluded only by a narrower lifecycle rubric. Replacing the reviewed decisions keeps the estimate positive (4.61 to 5.39 points) but moves its interval to or across zero. Re-grading all 64 reports with judges from two other model families, without arm labels or prior scores, agrees with the original judge on 91.8\% and 95.7\% of decisions (weighted $\kappa=0.64$ and $0.72$) and gives gains of 10.63 and 6.09 points. The study contributes a traceable evaluation that connects aggregate reward to contract-level evidence and judge sensitivity, together with concrete review checks for migration advice. Executable end-to-end repairs, independent human annotation, and other frameworks are left to future work.
\end{abstract}

\section{Introduction}
\begin{figure}[t]
\centering
\begin{tikzpicture}[
  font=\small,
  panel/.style={draw=gray!60, rounded corners=2pt, align=left, inner sep=6pt, text width=4.1cm, minimum height=3.4cm, anchor=north west},
  arr/.style={-{Stealth[length=2.2mm]}, very thick, gray!80},
]
\node[panel] (task) at (0,0) {\textbf{Task S11 (static diagnosis)}\\[3pt] Extracted archive paths must stay \emph{inside} the plugin directory.};
\node[panel, right=0.8cm of task.north east, anchor=north west] (answer) {\textbf{Archived answer (check, simplified)}\\[3pt] for non-absolute relative paths:\\reject if prefixed by \texttt{..} + separator\\[4pt] input \texttt{..} $\Rightarrow$ \textcolor{vermillion}{\textbf{accepted} $\times$}};
\node[panel, right=0.8cm of answer.north east, anchor=north west] (grade) {\textbf{Grading outcome}\\[3pt] original judge: full containment credit \textcolor{vermillion}{$\times$}\\[4pt] bounded review: partial credit\\parent escape confirmed by probe};
\draw[arr] (task.east) -- (answer.west);
\draw[arr] (answer.east) -- (grade.west);
\end{tikzpicture}
\caption{A proposed containment guard accepts the parent directory despite full original containment credit. The archived with-skill answer to task S11 proposes a containment predicate that accepts the literal path \texttt{..}, the parent directory, and the original judge still awards full containment credit; bounded review lowers the criterion to partial credit. Sections~\ref{sec:mechanisms} and~\ref{sec:rq2} give the full analysis.}
\label{fig:motivating}
\end{figure}
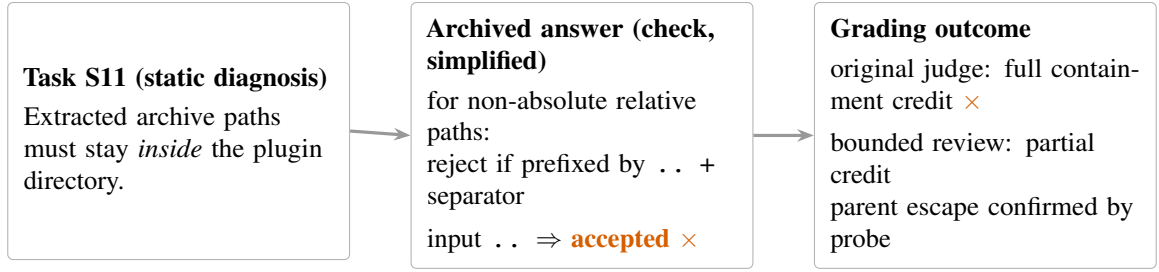

Software migration requires more than replacing names in source code. API changes routinely break client code~\citep{dig2006apievolve,raemaekers2017semver}, and clients often defer updates because migration is costly~\citep{kula2018update}. A repair must respect the contract of the target version, distinguish public APIs from internal implementation, and preserve runtime behavior. Plugin systems make these requirements particularly visible: a change can affect loading, authentication, rendering, event dispatch, or process shutdown independently. Advice that sounds consistent with a release note may still recommend an unavailable API or omit a boundary condition. In one archived answer studied below, a proposed path-containment predicate accepts the parent directory, and an automated judge still awards it full containment credit (Figure~\ref{fig:motivating}).

Large language models (LLMs) now support a wide range of software-engineering tasks~\citep{fan2023llm4se,hou2024llm4se}, and coding agents resolve repository-level issues by inspecting, editing, and testing code~\citep{yang2024sweagent,wang2025openhands,xia2025agentless}. An agent skill bundles instructions, reference material, and sometimes executable tools that such an agent loads when relevant~\citep{anthropic2025agentskills}. Such a package can reduce the effort needed to locate version-specific facts. It can also introduce stale instructions, unnecessary work, or confidence in an incomplete repair. Existing skill benchmarks already demonstrate heterogeneous benefits and overhead~\citep{skillsbench,swe-skills-bench,webdev-skills-bench}. Consequently, the contribution of a maintenance case study cannot rest only on showing that a skill sometimes helps.

The practical question is whether an observed score gain identifies advice a maintainer can act on. An answer can name the correct migration concept while leaving the required operation unavailable, an edge case unhandled, or a runtime condition unmet. Aggregate reward does not show which of these distinctions was checked.

We investigate the shipped plugin-upgrade skill for \framework{}, linking task contracts to archived answers and the judgments that reward them. The available evidence accumulated during development rather than under one prospectively controlled experiment. We retain this history, distinguish incompatible protocols, and give greatest analytical attention to a later comparison with archived outputs, per-criterion judgments, and execution records. The study asks:
\begin{enumerate}
\item RQ1: What changes in recorded reward accompany skill availability in the focal comparison, and how are they distributed across tasks?
\item RQ2: Where does recorded contract credit change, and what do inspected answers show about the adequacy of migration advice and its grading?
\item RQ3: How do resource accounting and grading sensitivity qualify the apparent benefit?
\end{enumerate}

The study makes three contributions.
\begin{itemize}
\item An auditable within-configuration comparison that links 64 answers to version-pinned tasks, original criterion judgments, and resource records (Sections~\ref{sec:setting} and~\ref{sec:focal-method}).
\item A two-level account of grading: complete contract-stratified accounting of all 328 decisions, a bounded review that locates judgments missing an operational requirement, and a blinded cross-family re-grading that measures how much the result depends on the judge (Sections~\ref{sec:bounded_review}--\ref{sec:contracts}, \ref{sec:rq2}, and~\ref{sec:rq3}).
\item Executable predicate and timer checks that separate an observed defect from an alternative repair excluded by a narrower rubric, and three review questions for maintainers derived from these cases (Sections~\ref{sec:mechanisms} and~\ref{sec:review_questions}).
\end{itemize}
The findings concern one plugin framework and static diagnosis; Section~\ref{sec:threats} discusses validity and Section~\ref{sec:future} the studies that would extend them.

\section{Related Work}\label{sec:related}
\paragraph{Skills and software engineering.}
SkillsBench evaluates skills across diverse tasks~\citep{skillsbench}. SWE-Skills-Bench examines their utility in software engineering~\citep{swe-skills-bench}. SkillLens studies skill generation and consumption~\citep{skilllens}, while WebDev-Skills-Bench compares skill conditions in web development, including controls for supplied context~\citep{webdev-skills-bench}. These studies establish that skill benefits are heterogeneous and carry resource overhead. We complement their breadth with depth: an artifact-traceable account of version-contract errors and evaluation sensitivity in one maintained plugin-migration setting.

\paragraph{API evolution and migration.}
Breaking API changes are frequent and costly for clients. Most client-breaking changes in the frameworks studied by \citet{dig2006apievolve} are refactorings; about a third of Maven releases, including minor ones, introduce at least one breaking change~\citep{raemaekers2017semver}; and ecosystems differ in who bears the cost of a break~\citep{bogart2016break}. Library developers break APIs mainly to add features and simplify interfaces~\citep{brito2018whyhow}, whereas clients lag behind API updates~\citep{mcdonnell2013android,kula2018update}, and dependency constraints only partly follow semantic versioning~\citep{decan2021semver}. \citet{lamothe2021apievolution} survey this literature. Migration support has progressed from edits mined from migration histories~\citep{xu2019meditor} to LLM-based migration, from first library-migration results~\citep{almeida2024llmmigration} to industrial migrations at Google~\citep{ziftci2025migrating,nikolov2025google}. VersiCode and CODEMENV benchmark version-controllable code generation and code migration~\citep{versicode,codemenv}. These studies motivate our focus on version contracts: the target of a migration is defined by what the new version exposes and requires, not by the names that changed.

\paragraph{Coding agents and their evaluation.}
Surveys document the rapid adoption of LLMs across software-engineering tasks~\citep{fan2023llm4se,hou2024llm4se}. Agent frameworks such as SWE-agent and OpenHands give models tools for navigating and editing repositories~\citep{yang2024sweagent,wang2025openhands}, and a simpler localize-repair-validate pipeline is competitive with such agents on SWE-bench Lite~\citep{xia2025agentless}. Evaluations such as SWE-bench and AgentBench emphasize task and environment design~\citep{swe-bench,agentbench}. Our setting combines migration diagnosis with historical executable tasks; the most completely archived comparison consists of static answers, so its reward measures diagnostic advice rather than executable repair success.

\paragraph{LLM judges and measurement validity.}
LLM judges enable scalable evaluation~\citep{mt-bench-judge,chiang2023llmeval,liu2023geval}, including of code~\citep{zhuo2024icescore}, and their agreement with human scores has been measured for code translation, generation, and summarization~\citep{wang2025llmjudgese}. They exhibit position bias~\citep{wang2024notfair} and favor their own generations~\citep{self-preference}, which matters when solvers and judges share a model family. Recent surveys and a large-scale judge study distinguish agreement and consistency from validity~\citep{gu2024judgesurvey,he2025courtroom,judge-validity}. A panel of judges from different model families reduces single-model bias~\citep{verga2024juries}; our cross-family re-grading applies this idea as a sensitivity analysis (Section~\ref{sec:panel}). For empirical software-engineering studies involving LLMs, contamination, non-determinism, and closed models are recognized threats~\citep{sallou2024breaking,wagner2025towards}. Development exposure is one form of contamination: repeated use of the same task during skill refinement does not create a fresh holdout~\citep{time-travel-contamination}. We record each task's exposure (Section~\ref{sec:exposure}).

\section{Setting and Evidence}\label{sec:setting}
\subsection{Versioned migration and the intervention}\label{sec:intervention}
\framework{} is a plugin-based framework with a marketplace and frequent releases. Relevant changes include renamed exports, authentication boundaries, build-artifact layouts, session representations, and version-routing rules. Tasks specify an initial state, a target contract, permitted operations, and a deliverable. Static tasks request diagnosis; hands-on tasks require a repair in a version-pinned environment. A host cold boot checks liveness, while task-specific probes exercise additional requirements. Neither establishes correctness beyond the checked behaviors.

Upgrade cards record symptoms, causes, actions, version applicability, and provenance. For example, card \texttt{DSH-0.1.2-A1-01} concerns removal of the Host APIProxy surface on the 0.1.1-rc.2 to 0.1.2-alpha.1 edge. Correct migration requires locating actual call sites and selecting the corresponding Remote interfaces, rather than applying a client-plane recipe to every plugin. This illustrates why merely mentioning the correct release or card is insufficient.

The shipped skill links inspection, version-corridor analysis, implementation, and verification. Cards can inform both the skill and the benchmark. Thus the comparison is an open-book evaluation of the complete package, including access to facts, rather than an isolated test of procedural reasoning. Historical repository states differ, so the intervention is not assumed byte-identical across configurations.

\subsection{Task construction and exposure}\label{sec:exposure}
Contributors turn migration incidents and version changes into cards, fixtures, instructions, reference solutions, and graders. Contributor self-check and maintainer review constitute development quality control, not independent blinded annotation. The repository's exposure ledger distinguishes feedback used to refine the skill, grader development, and uncleared exposure. We make no claim that the evaluated tasks form an independent holdout.

An immutable earlier snapshot contains \BenchmarkTaskCount{} tasks (\BenchmarkStaticCount{} static and \BenchmarkHandsOnCount{} hands-on). Task identifiers carry a type prefix (S for static, M for mixed-mode, H for hands-on), and this snapshot contains S1--S8. The static pool was later extended to 22 tasks (S1--S22); the historical GLM configurations scored that pool in full, and the focal comparison draws its recorded selection of 16 static tasks from it. This inventory is not the denominator of every experiment. Historical configurations use 56, 23, 21, or 22 scored tasks. Repeated attempts are nested within tasks; they do not increase the number of independent migration problems. Related tasks may also share incident sources, limiting task-level uncertainty estimates.

\subsection{Evidence tiers}
We distinguish three evidence tiers. The focal comparison answers the research questions; the other tiers document the available context.
\begin{enumerate}
\item Focal archived comparison: 16 tasks, two conditions, and two repeats, with 64 reports, original verdicts, scores, selection and execution records, and an offline integrity audit.
\item Historical context: five within-configuration paired summaries, with heterogeneous task pools, aggregation, grading, and raw-artifact coverage.
\item Supplementary evidence: a newer 16-task Qwen comparison with scored records but without archived original answers and judgment reasons, plus three GLM-5.3 rounds with a change of judge in round three. Neither is pooled into the focal estimate.
\end{enumerate}

\section{Method}
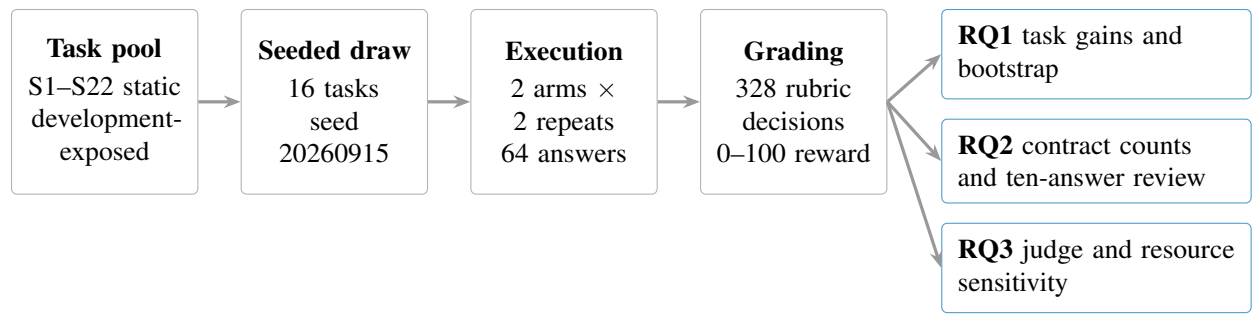
\begin{figure}[H]
\centering
\resizebox{\linewidth}{!}{%
\begin{tikzpicture}[
  font=\small,
  stage/.style={draw=gray!60, rounded corners=2pt, align=center, inner sep=6pt, text width=2.0cm, minimum height=2.4cm},
  analysis/.style={draw=oiblue!70, rounded corners=2pt, align=left, inner sep=6pt, text width=3.6cm},
  arr/.style={-{Stealth[length=2.2mm]}, very thick, gray!80},
]
\node[stage] (pool) at (0,0) {\textbf{Task pool}\\[2pt] S1--S22 static\\development-exposed};
\node[stage, right=0.55cm of pool] (sel) {\textbf{Seeded draw}\\[2pt] 16 tasks\\seed 20260915};
\node[stage, right=0.55cm of sel] (exec) {\textbf{Execution}\\[2pt] 2 arms $\times$ 2 repeats\\64 answers};
\node[stage, right=0.55cm of exec] (judge) {\textbf{Grading}\\[2pt] 328 rubric decisions\\0--100 reward};
\node[analysis, anchor=north west] (rq1) at ($(judge.north east)+(0.7,0)$) {\textbf{RQ1} task gains and bootstrap};
\node[analysis, below=0.25cm of rq1] (rq2) {\textbf{RQ2} contract counts and ten-answer review};
\node[analysis, below=0.25cm of rq2] (rq3) {\textbf{RQ3} judge and resource sensitivity};
\draw[arr] (pool) -- (sel);
\draw[arr] (sel) -- (exec);
\draw[arr] (exec) -- (judge);
\draw[arr] (judge.east) -- (rq1.west);
\draw[arr] (judge.east) -- (rq2.west);
\draw[arr] (judge.east) -- (rq3.west);
\end{tikzpicture}
}
\caption{Overview of the focal study design. A seeded stratified draw selects 16 of the 22 static tasks; two arms with two repeats produce 64 archived answers, graded by the report-judge-v2 rubric into 328 criterion decisions. Three analyses consume this archive: the focal paired estimand (RQ1), complete contract-stratified accounting and a bounded ten-answer review (RQ2), and grading-sensitivity plus resource accounting (RQ3).}
\label{fig:overview}
\end{figure}

We report the study as a retrospective single-case study~\citep{runeson2009case}; in the terms of \citet{stol2018abc}, it is a field study that favors realism over generalizability, such studies are covered by the case-study standard of the ACM SIGSOFT empirical standards~\citep{ralph2020empirical}. Figure~\ref{fig:overview} summarizes the focal study design end to end: a seeded selection from the static task pool, paired execution under two conditions, criterion-level grading, and the three analyses that answer the research questions. The subsections below follow this flow.

\subsection{Focal comparison and estimand}\label{sec:focal-method}
The focal run is the archived GLM-5.3-Flash unified S16 comparison; ``S16'' here names the archived configuration of sixteen selected static tasks, not the task identifier S16. The model label is the recorded gateway label, not an independently verified served-weight identity. Each task has two no-skill and two with-skill outputs, run as subagent sessions under the ZCode agent host. The conditions share a prompt template, but workspace paths differ. Two focal tasks (S5 and S6) carried run-time brief wording that described the diagnosis as closed-book, which did not anticipate a mounted skill; the briefs have since been amended to acknowledge harness-mounted materials, and the archived task packets preserve the run-time text under hash.

The selection script draws 16 of the 22 static tasks S1--S22 using seed 20260915, stratifying by problem family and fixture-size tercile with proportional quotas. Balance constraints require at least two hard tasks, two Chinese-prompt tasks, and two large-fixture tasks. The selector reads metadata and fixture sizes rather than scores; it excludes tasks S3, S4, S7, S8, S10, and S16. The selected set contains five static-contract diagnosis, seven runtime/client API, three release/install, and one Cordis profile task (Cordis is the underlying plugin-runtime dependency of \framework{}); 14 prompts are English and two are Chinese. The draw is reproducible, and it is drawn from the development-exposed pool described in Section~\ref{sec:exposure}.

The schedule alternates arm order in task blocks and reverses it in round two; the recorded execution order is retained. Four pilot cells on excluded tasks S4 and S10 are not included in the 64 formal cells. Formal execution used no hard wall-time cap, and all 64 cells produced scored reports.

Let $y_{tar}$ be the original score for task $t$, arm $a$, and repeat $r$. We calculate $\bar y_{ta}=(y_{ta1}+y_{ta2})/2$, $d_t=\bar y_{t,\mathrm{skill}}-\bar y_{t,\mathrm{no}}$, and $\widehat\Delta=16^{-1}\sum_t d_t$. Thus tasks receive equal weight. The focal percentile bootstrap~\citep{efron1993bootstrap} resamples paired tasks 10,000 times (mulberry32, a small deterministic pseudo-random number generator, seed 20260915). It quantifies variability over the observed tasks under resampling assumptions, not uncertainty over arbitrary repositories or model versions.

The report-judge-v2 rubric, the harness's criterion-level grading protocol, assigns a weight to each criterion. A pass receives its full weight, partial receives half, and fail or missing receives zero; declared triggered caps can limit the total. Criteria and frozen reference excerpts are supplied to the judge, which returns decisions and reasons rather than a total. The deterministic scorer calculates the 0--100 reward. Proposed verification can satisfy a static task; describing a test does not establish that it was executed.

The same model family supplies semantic evaluation. The run records describe arm-blind staging with one report per judgment; our later review was not blind. The archive retains original verdicts, including historical transport metadata; corrected provenance records distinguish harness-declared judge identity from independently observed provider identity. Scores are recomputed from rubric decisions. An integrity audit checks all 64 report hashes and all 328 original criterion decisions. Two stored reports require reversal of documented relative-link rewriting to reproduce their original hashes; this transformation is explicit rather than silently ignored.

\subsection{Bounded output review}\label{sec:bounded_review}
The initial targeted review was AI-assisted and non-blind, separate from the original verdicts. An earlier purposive pass covered five complete answers selected for prominent effects or operational concerns. A subsequent selection was committed before reading the additional full reports: paired answers for a positive task (S6), a negative task (S18), and a zero-change task (S12), at fixed repeats. One answer overlapped with the earlier pass. The combined scope is ten unique answers and 56 criteria.

The initial reviewer had access to task rubrics, reports, and earlier judgments, and selection used known outcomes. Multiple contributing plugin authors, who are domain experts for these plugins rather than independent annotators, subsequently performed the human checking and reported broadly similar assessments; the later repository revision identifies them as non-blind reviewers. The ten archived answers retain the same 56 decisions and replacement scores across these revisions. No separate per-reviewer scoring sheet or item-level confirmation was supplied, so revision stability is not interpreted as measured human--AI agreement or independent blinded validation.

The provenance ledger retains both the initial AI-assisted analysis and the reported human follow-up. Selection, report hashes, criterion-level reasons, and replacement scores are recorded. We keep original results as the primary recorded endpoint and show both the earlier and the combined replacement sensitivities. Replacements affect only reviewed answers; unreviewed scores are not presumed correct.

\subsection{Cross-family judge panel}\label{sec:panel}
Because the focal solver and judge share a model family, we re-judged all 64 focal reports with judges from other families under a protocol and analysis plan written before any panel verdict existed (Appendix~\ref{app:panel}). Each item contains the unchanged report-judge-v2 system prompt and judge input (task, rubric, frozen excerpts, citation audit, and report), and nothing about arm, repeat, original verdicts, or review outcomes; item identifiers are opaque hashes and the unblinding key is hash-pinned.

Claude Opus 5.5 (\texttt{claude-opus-5-5}) ran as 16 fresh agent sessions of four items each, so that no session saw two answers to the same task. GPT-5.5 (\texttt{openai-gpt-5.5}) processed the 64 items sequentially in one isolated Codex CLI session. Neither judge saw the other's verdicts. Appendix~\ref{app:panel} records session provenance.

Blinding is limited to supplied labels and prior outcomes. Arm is only partly concealed: 18 of 32 with-skill reports mention the skill in their own text, as they did for the original judge. The committed reports include the documented relative-link rewriting noted above.

Following the analysis plan, we measured criterion-level agreement on the credit scale (pass, partial, fail or missing) with linearly weighted Cohen's $\kappa$~\citep{cohen1960coefficient,cohen1968weighted} and recomputed the focal estimand under each judge and under the per-answer judge mean. Agreement with the author review is reported as a reference point; that review started from the original decisions and is not ground truth. The panel measures sensitivity to judge configuration; human validation is discussed in Section~\ref{sec:future}.

\subsection{Complete contract-stratified analysis}\label{sec:contracts}
We map each of the focal tasks' existing rubric criteria to one primary domain: version/release applicability; API/data/ownership; lifecycle/ordering/deployment; safety/boundary handling; failure attribution; or evidence/verification/provenance. The mapping is retrospective and fixed before generating the domain table. Multi-part criteria receive one primary label to avoid double counting, so these categories are an analytical organization of the rubric rather than an independently validated failure taxonomy.

Every formal report is linked to its task packet and original criterion verdicts through stored hashes. The analysis accounts for all 64 reports and all 328 criterion decisions, including original successes and shortfalls. We report full-credit counts with per-arm criterion denominators and distinct task counts. The counts are recorded judge outcomes; semantic re-grading is the role of the bounded review and the judge panel. Repeats are nested within tasks, tasks can contribute to multiple domains, and we make no domain-level significance claims.

\subsection{Bounded mechanism and robustness checks}\label{sec:mechanisms}
After case selection and the original analysis, we added offline checks under a recorded retrospective protocol. For S11, we extracted the archived proposed containment function, removing only TypeScript annotations, and exercised seven declared lexical inputs under each of Node's POSIX and Windows path algorithms. Parent and ancestor escapes, ordinary and dotted descendants, prefix siblings, and platform-specific controls distinguish the observed defect from ordinary valid behavior. The probe exercises the predicate itself, not an HTTP route, native Windows filesystem operations, or symlink resolution.

For S18, the fixture describes a timer planner but does not contain its implementation. We therefore reconstructed a minimal self-rearming timer in separate Node child processes. Five conditions compare referenced timers, cleanup that is not invoked, explicit unmount with cancellation, unreferencing every timeout, and unreferencing only the first timeout while another handle initially keeps the process alive. Each child emits a readiness marker; a parent watchdog terminates children still alive 700 milliseconds later. Natural exit is distinguished from watchdog termination. These are mechanism checks with controls, not executions of the original plugin or new scored solver attempts.

For the original and both existing replacement endpoints, we also enumerate all $2^8=256$ sign assignments of the eight nonzero paired task differences. The two-sided tail fraction counts assignments whose absolute sum is at least the observed absolute sum, including ties. This is a descriptive sensitivity under independent, sign-exchangeable task differences; the deterministic arm schedule does not justify a causal randomization test. We additionally report every leave-one-task-out mean and both repeat-wise means. None of these endpoints is selected by its significance.

\section{Results}
\subsection{RQ1: Recorded reward changes}
\begin{table}[t]
\centering\small
\begin{tabular}{lrrrl}
\toprule
Focal endpoint & No skill & Skill & $\Delta$ & 95\% task-bootstrap interval\\
\midrule
Original judgments &93.83&98.75&4.92&[0.31, 10.86]\\
Earlier S11 replacement &93.83&98.44&4.61&[$-0.23$, 10.63]\\
All reviewed replacements &93.05&98.44&5.39&[0.00, 11.80]\\
\bottomrule
\end{tabular}
\caption{Focal 16-task comparison. Original scores remain unchanged in the archive. Replacement rows are partial-review sensitivity analyses, not validated estimates from a fully regraded dataset.}
\label{tab:focal}
\end{table}

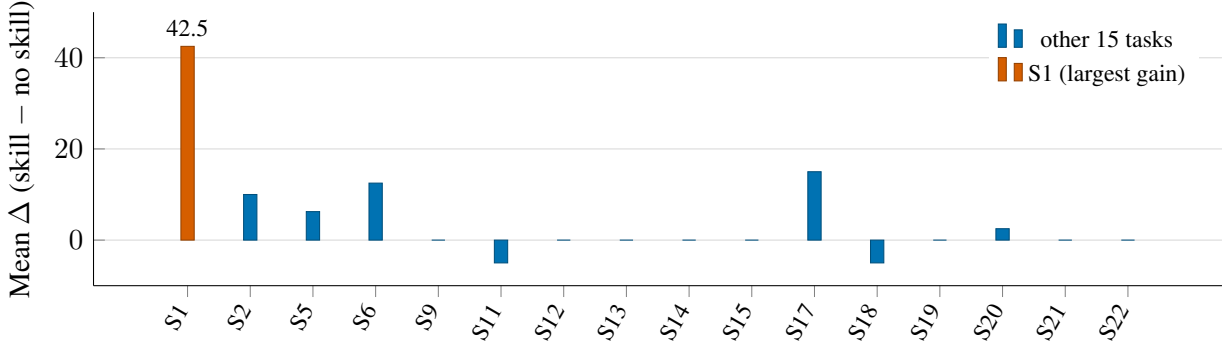
\begin{figure}[t]
\centering
\begin{tikzpicture}
\begin{axis}[
  ybar, bar width=5pt, bar shift=0pt,
  width=\columnwidth, height=5.2cm,
  symbolic x coords={S1,S2,S5,S6,S9,S11,S12,S13,S14,S15,S17,S18,S19,S20,S21,S22},
  xtick={S1,S2,S5,S6,S9,S11,S12,S13,S14,S15,S17,S18,S19,S20,S21,S22},
  x tick label style={rotate=60, anchor=east, font=\footnotesize},
  ylabel={Mean $\Delta$ (skill $-$ no skill)},
  ymin=-10, ymax=50,
  ymajorgrids, grid style={gray!30},
  axis y line*=left, axis x line*=bottom,
  legend style={font=\footnotesize, at={(0.98,0.98)}, anchor=north east, draw=none},
]
\addplot+[fill=oiblue, draw=oiblue!70!black] coordinates {(S2,10) (S5,6.25) (S6,12.5) (S9,0) (S11,-5) (S12,0) (S13,0) (S14,0) (S15,0) (S17,15) (S18,-5) (S19,0) (S20,2.5) (S21,0) (S22,0)};
\addplot+[fill=vermillion, draw=vermillion!70!black] coordinates {(S1,42.5)};
\legend{other 15 tasks, S1 (largest gain)}
\node[font=\footnotesize,anchor=south] at (axis cs:S1,42.5) {42.5};
\end{axis}
\end{tikzpicture}
\caption{Task-level mean reward changes in the focal comparison are concentrated in one task and bounded by ceiling effects. S1 alone gains 42.5 points, eight task pairs score 100 in both arms (zero change), and two tasks decline by 5 points. Per-task scores are in Table~\ref{tab:focal_tasks}.}
\label{fig:task_deltas}
\end{figure}

In the focal comparison, mean original reward increases from 93.83 to 98.75 (Table~\ref{tab:focal}). Six tasks improve, two decline, and eight have zero change (Figure~\ref{fig:task_deltas}); all eight zero-change pairs are at 100 in both arms. S1 has a 42.5-point task-level gain. Removing it reduces the overall gain to 2.42 points across the remaining 15 tasks. The original Wilcoxon signed-rank approximation~\citep{wilcoxon1945individual} gives $p=0.0797$ with eight nonzero pairs. The corresponding exhaustive sign-enumeration tail fraction is $20/256=0.0781$. Repeat-wise mean gains are 4.53 and 5.31; all 16 leave-one-task-out means range from 2.42 to 5.58. Every variant keeps the direction positive. Because eight pairs are at the ceiling and one task contributes most of the gain, we read the focal estimate as a descriptive association.

The focal estimate describes this configuration and task set. Historical and supplementary configurations, which use different task pools, hosts, and judges, are reported separately in Appendix~\ref{app:historical}.

\subsection{RQ2: Contract failures and grading disagreements}\label{sec:rq2}
\input{generated/contract-coverage-table.tex}
\input{generated/contract-credit-figure.tex}
Figure~\ref{fig:contract_credit} displays full-credit proportions across the full cohort; Table~\ref{tab:contracts} gives the counts and task coverage. Original full-credit counts rise from 8 to 12 out of 12 version-related decisions and from 35 to 45 out of 46 API/data/ownership decisions per arm. Failure attribution has full credit in both arms (32/32), as does the safety/boundary domain (10/10). These counts describe the original judge; the reviewed S11 containment defect demonstrates why an all-full-credit domain cannot be called verified safe. Domains contain different tasks and weights, so their percentages are not a difficulty ranking or an estimate of a causal skill mechanism.

The bounded targeted review finds three criterion disagreements in two of ten answers. The sample is purposive, so these disagreements identify concrete weaknesses in the recorded measurement rather than an error rate.

\paragraph{Information access and public API boundaries.}
S1's reviewed answers retain their original 55 and 100 scores. Part of its rubric rewards mapping to supplied migration cards. The large gain therefore includes access to organized facts and identifiers, and cannot isolate a procedural reasoning benefit. In S6, the no-skill answer correctly recommends removing obsolete defensive code but incorrectly attributes informational-event handling wholly to the host. It also recognizes a public append API gap while recommending an unqualified append route. Review changes two criteria and lowers this answer from 87.5 to 62.5. The paired skill answer retains 100 by distinguishing producer responsibilities, retention, and unavailable public surfaces. The resulting sensitivity increases estimated skill benefit, illustrating that grading errors can favor either arm. Additional full-text inspection of S6 repeat 2 finds the same platform-only interpretation in the no-skill answer, whereas both skill answers distinguish retention from a supported producer seam. This is repetition within one task, not a second independent incident; no additional score replacement is imposed in the fixed ten-answer sensitivity.

\paragraph{Containment boundaries.}
In S11, a with-skill answer proposes a relative-path check equivalent to accepting an empty relative path or any non-absolute path that does not start with two dots followed by a separator. The exact relative path \texttt{..} passes this predicate, even though it denotes the parent directory. The extracted predicate accepts the parent in both path algorithms, contradicting the declared containment expectation in two of 14 diagnostic inputs; the other 12 control inputs behave as expected. Review lowers the containment criterion from full to partial credit and the answer from 90 to 80. This confirms a defect in the proposed guard; whether it is exploitable depends on the surrounding route, which the probe does not execute.

\paragraph{Cleanup versus process liveness.}
For S18 repeat 2, the no-skill answer retains 100 and the skill answer retains 90. The skill answer proposes explicit host teardown, while the rubric requires timers that do not keep a still-mounted probe host alive. Our child-process reconstruction makes the distinction observable: a referenced chain and an uninvoked cleanup callback remain alive until the watchdog stops them; both explicit unmount with cancellation and unreferencing every timeout exit naturally. Unreferencing only the first timeout is insufficient when an initially live companion handle allows a referenced successor to be scheduled.

Thus explicit teardown is a functioning alternative in the reconstruction when the host performs it. The retained partial credit measures noncompliance with the rubric's narrower mounted-host requirement, not proof that the proposed cleanup can never repair a CI hang. The instruction asks for a one-line hang fix, whereas the rubric specifies timer unreferencing. This difference exposes sensitivity to how acceptable repairs are defined. We preserve the original and prior reviewed scores; the new mechanism check does not silently create another replacement endpoint.

\paragraph{Ceiling and a retained partial score.}
Both reviewed S12 answers retain 100: they connect a Windows file lock to the native host process, distinguish the recorded release channels, and prescribe the pinned installation and restart sequence. This zero-change pair illustrates a ceiling case rather than evidence that the skill was consumed or unnecessary in all contexts. The previously reviewed S17 no-skill answer retains 80 because recognizing onboarding requirements does not repair an incorrect injection descriptor.

\subsection{RQ3: Sensitivity and resource use}\label{sec:rq3}
\begin{figure}[t]
\centering
\resizebox{\linewidth}{!}{%
\begin{tikzpicture}
\begin{axis}[
  width=\columnwidth, height=4.6cm,
  xlabel={Mean paired $\Delta$ (skill $-$ no skill)},
  symbolic y coords={All reviewed replacements, Earlier S11 replacement, Original judgments},
  ytick=data, y tick label style={font=\small},
  xmin=-1.5, xmax=13,
  xmajorgrids, grid style={gray!30},
  axis y line*=left, axis x line*=bottom,
]
\addplot[gray, dashed, no markers, forget plot] coordinates {(0,Original judgments) (0,All reviewed replacements)};
\addplot+[only marks, mark=*, mark size=1.8pt, color=oiblue, error bars/.cd, x dir=both, x explicit, error bar style={thick}]
coordinates {
  (4.92,Original judgments) +- (4.61,5.94)
  (4.61,Earlier S11 replacement) +- (4.84,6.02)
  (5.39,All reviewed replacements) +- (5.39,6.41)
};
\end{axis}
\end{tikzpicture}
}
\caption{Grading sensitivity of the focal gain. Replacing reviewed criterion decisions moves the mean change between 4.61 and 5.39 points; the earlier S11-replacement interval crosses zero and the all-reviewed interval touches it (dashed line), so the positive direction survives while the strength of evidence depends on grading decisions. Whiskers are 95\% task-bootstrap intervals.}
\label{fig:sensitivity}
\end{figure}

Replacing only the reviewed S11 score changes the focal estimate to 4.61 points and moves its interval across zero (Figure~\ref{fig:sensitivity}). Replacing all reviewed scores changes it to 5.39 points with a lower endpoint at zero. The direction remains positive, but the strength of evidence depends on grading decisions. The exhaustive sign-enumeration tail fractions are $34/256=0.1328$ for the S11-only replacement and $30/256=0.1172$ for all reviewed replacements, compared with 0.0781 originally. Neither the larger replacement estimate nor the original positive interval should be selected in isolation.

Figure~\ref{fig:judge_sensitivity} compares complete judge re-grades and their per-report means; Table~\ref{tab:judge_panel} gives the exact estimates and agreement statistics.

\input{generated/judge-sensitivity-figure.tex}
\input{generated/llm-judge-panel-table.tex}

Claude agrees with the original GLM judge on 301 of 328 criterion credit levels (91.8\%, weighted $\kappa=0.64$); GPT agrees on 314 (95.7\%, $\kappa=0.72$). Claude and GPT agree with each other on 297 (90.5\%, $\kappa=0.57$). In the conventional bands of \citet{landis1977measurement}, agreement with the original judge is substantial for both panel judges and moderate between them. Their answer scores match the original in 45 and 53 of 64 cases, respectively (Table~\ref{tab:judge_panel}).

The judges differ in both direction and extent of re-scoring. Claude assigns lower credit than GLM to 26 decisions and higher credit to one, with 23 of those reductions on no-skill answers. GPT's departures are smaller and mixed: five lower-credit and nine higher-credit decisions. Re-graded gains are 10.63 [3.44, 19.14] for Claude and 6.09 [1.56, 11.09] for GPT, versus the original 4.92 [0.31, 10.86]. The per-answer mean across all three judges gives 7.21 [1.98, 13.46]; the mean of the two cross-family judges gives 8.36 [2.66, 14.96].

Agreement does not establish accuracy. The original judge awards full credit to 300 of 328 decisions (91.5\%), so high exact agreement must be read alongside the chance-adjusted weighted $\kappa$. Against the 56 decisions in the non-blind author review, GLM, Claude, and GPT match 53, 51, and 48; that review was seeded by GLM's decisions and is no gold standard. Both cross-family re-judgments preserve a positive descriptive direction, but the magnitude varies, and the judges share a rubric whose validity is not established by their agreement. The original judgment remains the primary recorded endpoint.

The 64 formal execution records contain 3,185,993 no-skill and 12,320,379 skill \texttt{subagent\_tokens}, a ratio of 3.87. The corresponding sums of recorded task durations are 9,446 and 11,450 seconds, a ratio of 1.21. The token field does not separately define input, output, and cache accounting, so it is not a billing-cost measure. Summed task durations are not end-to-end wall-clock time under overlapping execution. Pilot records are excluded.

Together, these results answer RQ3 with a conditional assessment: higher recorded reward accompanies greater recorded token use in the focal configuration. The bounded review's replacement intervals touch or cross zero, whereas both label-blinded LLM re-judgments retain positive intervals under the specified task bootstrap. These analyses probe different sources of grading sensitivity and do not establish a single corrected score. The archive supports neither a monetary cost-benefit estimate nor a claim that the added resource use produces more successful repairs.

\section{Discussion}
\subsection{What contract inspection adds to aggregate reward}
The focal result has two complementary interpretations. Skill availability is associated with a higher mean recorded score, with much of the gain concentrated in one task and no room for improvement on eight others. Inspection then asks whether the rewarded advice meets the particular requirement that makes a migration usable. This second question matters even when the average improves: S11 receives full containment credit for a predicate that accepts the parent directory, whereas S18's partial credit depends on requiring exit without host teardown. The mechanism check shows that teardown can also permit exit when performed. The cases distinguish a missed predicate defect from a narrower choice of acceptable lifecycle repair.

Full-cohort accounting and targeted inspection answer different questions. The former shows where the original judge assigned credit across every focal answer, avoiding a results table built solely from selected successes or failures. The latter supplies counterexamples with reasons, but cannot estimate how often such defects occur. For example, the safety/boundary domain has full original credit in both arms even though the S11 check fails a concrete boundary case. Neither the domain count nor the selected defect alone characterizes the reliability of the whole dataset.

\subsection{Review questions grounded in the observed cases}\label{sec:review_questions}
The cases motivate three checks for maintainers reviewing migration advice. For each proposed API action, verify that the target version exposes a callable public surface and identify the component responsible for producing or retaining the data (S6). For a boundary predicate, test the exact boundary value as well as ordinary descendants; the parent path itself supplies a counterexample in S11. For lifecycle advice, state the required runtime condition separately from its suggested cleanup action; releasing a component resource and allowing process exit are different requirements (S18).

These checks can be applied to an individual answer without relying on its total score. Their evidence differs in kind: the S11 predicate was executed directly, S18 was examined through a minimal timer reconstruction with positive and negative controls, and S6 is a static assessment against the recorded contract. They are review questions transferable to other projects; how often the underlying defects occur there, and whether building the checks into an agent or grader improves outcomes, are open questions (Section~\ref{sec:future}).

\subsection{Implications for evaluating maintenance skills}
A maintenance-skill evaluation should state which endpoint it measures: locating applicable knowledge, supplying contract-consistent advice, or completing a verified repair. The focal archive measures graded static advice. S1's card-mapping reward also captures access to organized facts (Section~\ref{sec:rq2}). Reporting task-level changes and the contract behind each selected shortfall makes these distinctions visible without treating the rubric as a substitute for execution.

In this archive, retaining original verdicts alongside explicit alternative scores is necessary to interpret the result. Correcting the reviewed S6 judgment increases estimated skill benefit, while correcting S11 reduces it. Thus review should inspect both arms and preserve disagreements that strengthen or weaken the apparent effect. Our replacement analyses demonstrate this sensitivity locally. S18 additionally shows why a grader should state whether alternative repairs are acceptable or whether a particular lifecycle condition is mandatory.

The historical ordering does not establish a capability-response curve; model behavior does not always improve monotonically with scale~\citep{inverse-scaling}. Baseline reward shares mathematical components with the gain and is bounded by the same ceiling; model comparisons also mix infrastructure, graders, task exposure, and skill versions. The present contribution is an evidence-linked maintenance case study; claims about an inverted U, production reliability, or cost effectiveness would need dedicated designs and endpoints.

\section{Threats to Validity}\label{sec:threats}
We organize threats following \citet{wohlin2012experimentation} and consider the LLM-specific threats identified by \citet{sallou2024breaking} and \citet{wagner2025towards}.
\paragraph{Construct validity.}
Static diagnostic reward is not functional repair success. Some criteria reward card mappings or metadata. Keyword and semantic graders measure different things, and numeric normalization does not harmonize their constructs. The focal solver and judge share a model family. The two cross-family panel judges (Section~\ref{sec:panel}) preserve the positive direction but yield gains of 6.09 and 10.63 points, compared with the original 4.92; all three are language models sharing one prompt and rubric, so agreement does not validate the rubric itself. The single GPT judging session may carry information between sequential items, unlike the original separate calls. Family and session structure are confounded in this panel, and each configuration was judged once; test--retest reliability is unmeasured.

The initial non-blind AI-assisted review is useful for finding counterexamples but does not certify all scores. The human follow-up was performed by contributing plugin authors rather than independent annotators, without a separate item-level scoring dataset, precluding an inter-rater agreement estimate.

Attribution to the skill's organization remains unverified. No raw-document or generic-procedure control arm was run, so the comparison separates skill availability from absence but cannot attribute the benefit to the skill's procedural organization rather than its bundled facts (Section~\ref{sec:intervention}). Full-cohort contract tabulation inherits the original judge decisions and should not be confused with complete revalidation. The S18 instruction and rubric differ in how narrowly they specify the hang repair; compliance with the latter is not equivalent to the absence of other functioning repairs.

\paragraph{Internal validity.}
The data are retrospective, and contributors developed the skill and tasks in an overlapping workflow. Historical selections are not assumed random, and the focal seeded draw does not remove development exposure. Historical execution budgets, environment failures, and protocols differ. Original model labels rely on recorded infrastructure declarations. The review selection was frozen before additional full-output inspection, but after outcomes were known; it is not preregistration of the underlying experiment. The agent host did not isolate arms in separate containers, and workspace paths differ between arms.

\paragraph{Statistical conclusion validity.}
There are only 16 focal tasks and two repeats per arm, which is few for evaluating stochastic agents~\citep{arcuri2011practical}. Eight task pairs are at the ceiling. Task-level resampling treats tasks as independent even when incident sources may overlap. No family-clustered analysis is reported (Section~\ref{sec:future}). The selected review does not estimate population error, and partial replacements leave unreviewed scoring uncertainty unresolved. Intervals, signed-rank approximations, and leave-one-out results are exploratory descriptions of the recorded data rather than confirmatory tests of a universal effect.

\paragraph{External validity and reproducibility.}
The focal tasks concern one plugin framework and static reports. Historical task inventories and artifacts have unequal coverage. The newer Qwen outputs and grading reasons are unavailable in the repository, limiting replication of its endpoint. Live models and dependencies can change even with pinned task sources. The archive supports deterministic recalculation of the focal scores and reported sensitivities, not a guarantee of identical future model outputs. Mechanism probes are restricted to an extracted lexical predicate and a reconstructed timer; they do not execute complete agent-produced migrations, and the bounded timer observation is not a production liveness guarantee.

\section{Future Work}\label{sec:future}
The archive and tooling released with this study make several follow-up studies direct to run.

\paragraph{Independent human validation.} The repository already contains an outcome-blind protocol for grouping tasks by incident family with two independent annotators and third-party adjudication (\path{paper/audit/task-annotation-v1}). Executing it would support family-clustered uncertainty estimates. A blinded human re-grading of a random sample of criterion decisions would calibrate the LLM judges against human judgment and yield an inter-rater agreement estimate.

\paragraph{Attribution and executable endpoints.} Raw-document and generic-procedure control arms would separate the value of organized facts from that of the skill's procedure. Running agent-produced repairs end to end in version-pinned hosts would replace diagnostic reward with repair success, and would test whether the S11 and S18 mechanisms matter in deployed plugins.

\paragraph{Judge reliability.} Repeated judging of each configuration, with model family and session protocol varied independently, would measure test--retest reliability and separate the two factors confounded in the present panel.

\paragraph{Prospective replication and transfer.} A temporal-holdout split and its execution plan are specified in the repository (\path{benchmark/holdouts}); running it, and applying the same protocol to other plugin frameworks, would test whether the effect and the review questions of Section~\ref{sec:review_questions} transfer beyond this setting, and whether adding those questions to an agent or grader improves outcomes.

\section{Conclusion}
In an archived 16-task comparison, making a plugin-upgrade skill available raised recorded reward by 4.92 points, with the gain concentrated in one task and bounded by ceiling effects. The more consequential finding concerns how such gains should be read. API availability and responsibility, exact boundary coverage, and process liveness cannot be inferred from a plausible recommendation or an aggregate score; they have to be checked at the contract the advice is meant to satisfy. Complete accounting of all 328 decisions, a bounded review, and executable probes located a missed boundary defect and an alternative repair excluded by a narrower rubric. Two blinded judges from other model families preserved the positive direction of the gain while changing its size. By keeping contracts, answers, scores, and counterexamples linked, the study makes both the observed benefit and its open correctness questions inspectable, and provides a reusable basis for the human validation, executable repairs, and cross-project replication outlined above.

\section*{Acknowledgments}
We thank the other contributors to the open-source repository, including the GitHub users whiteicey, AdamPlatin123, and HuanLinOTO, for benchmark tasks, experiment runs, and migration cards.

\section*{Artifact Availability and Research Transparency}
The working repository is \url{https://github.com/oh-my-dsh/dsh-plugin-upgrade-skill}. The focal archive is under \path{benchmark/results/artifacts/2026-09-15-glm-5.3-flash-unified-s16}; the supplementary review is under \path{paper/audit/output-review-20260917}. Original outputs and scores are preserved separately from review verdicts. Complete contract accounting and human-follow-up provenance are stored under \path{paper/audit/contract-review-20260917}. The cross-family judge panel, its protocol, and all panel verdicts are under \path{paper/audit/llm-judge-panel-v1}. Offline mechanism probes and their execution environment are archived under \path{paper/audit/mechanism-checks-20260917}. The claim-to-evidence ledger identifies inputs and offline commands for regenerating the analyses. Availability is described per configuration; no universal raw-output completeness claim is made.

\section*{Use of AI Assistance}
AI tools (Codex and ChatGPT) assisted with source inspection, analysis code, initial targeted output review, and manuscript drafting. Non-blind human checking was subsequently performed by multiple contributing plugin authors. The archive preserves both stages and does not claim independently measured human agreement. In the cross-family judge panel, Claude Opus 5.5 (via Claude Code) and GPT-5.5 (via an isolated Codex CLI session) served as LLM judges; Claude Code also assisted with panel tooling and manuscript revision. Panel verdicts are LLM judgments and are reported as such.

\bibliographystyle{plainnat}
\bibliography{custom}
\appendix
\clearpage
\section{Task-level focal results}
\input{generated/focal-task-table.tex}

\section{Historical and supplementary context}\label{app:historical}
\subsection{Historical paired summaries}\label{sec:paired}
For each historical configuration, we use its recorded task reward on a 0--100 scale and preserve its original within-task aggregation. Qwen uses means of three scored attempts, DeepSeek and the two GLM groups use medians of three runs or rounds, and Terra uses one attempt. These numeric scales do not imply equivalent constructs: keyword grading, semantic judgment, and composite executable rewards differ. We do not pool configurations or regress their gains on their baseline scores as an independent capability measure.

Historical percentile intervals use 10,000 task-paired bootstrap samples with mulberry32 seed 20260907. The reported two-sided Wilcoxon signed-rank calculations exclude zero differences and use a tie-corrected normal approximation with continuity correction. These exploratory summaries can disagree near a boundary; neither is a substitute for validating the endpoint. No multiple-comparison-adjusted confirmatory family is claimed.

\input{generated/paired-effect-table.tex}
Historical mean changes range from $-3.07$ to $+10.67$ points (Table~\ref{tab:paired_effect}). The older Qwen configuration has substantial skill-arm timeouts under its local infrastructure. Terra excludes H8 because both arms have verifier timeouts. The GLM groups mix semantic and keyword judgments. These differences prohibit attributing cross-row variation to model capability alone. A contaminated Luna baseline, which accessed a native plugin-creator skill on H2 and H3, is excluded from this five-configuration comparison.

Within the historical GLM pair, the Flash-minus-5.2 difference in mean task lift is 6.23 points, with a paired bootstrap interval [2.59, 10.05]. The round-specific Flash gains are 6.32, 8.82, and 9.82, versus 1.55, 1.64, and 5.00 for GLM-5.2. Leaving one task out yields lift gaps from 5.33 to 7.00. These completed robustness calculations support a descriptive contrast within the recorded protocols. Different judge mixes and development histories remain confounders; the contrast does not identify a capability-dependent response curve.

\subsection{Newer supplementary configurations}
A supplementary newer Qwen run on the same S16 configuration reports 71.17 to 77.89, a gain of 6.72 points, with a reported interval [$-12.42$, 26.25]. It contains 64 scored attempts, including five timeouts and two other nonzero-exit attempts. Termination status is distinct from the assigned score. Its raw answers and grading reasons are not archived, and the reported bootstrap randomization is insufficiently specified for exact interval reproduction. Fourteen tasks overlap with the focal Flash run; different hosts, budgets, judges, and selections prevent a controlled cross-model comparison. We use this result only as supplementary context.

Three supplementary GLM-5.3 rounds on S1--S22 are now archived. Recomputing all 132 verdicts against the pinned rubric, retaining half-point values before summation, gives round gains of 1.93, 3.18, and 6.59 points. The per-task median across rounds is 97.05 without the skill versus 98.64 with it, a gain of 1.59 points; its descriptive task-bootstrap interval is [$-1.36$, 5.80] (10,000 samples, seed 20260907). The third round used GLM-5.3 as judge rather than the GLM-5.3-Flash judge used in rounds one and two. The median cannot remove this protocol change, so these results remain a mixed-judge descriptive supplement.

Round three's S17 no-skill answer receives zero because its affirmative immediate-registration recommendation triggers the rubric's contradiction cap. The criterion decisions are retained: this pair contributes 100 of the round's 145 total gain points. Excluding S17 descriptively leaves a 2.14-point mean gain across 21 tasks; this is a sensitivity, not a basis for removing the observation. The archive discloses quota-related relaunches, concurrency changes, and a truncated round-one answer. These rounds do not establish a capability ladder or restore comparability with older protocols.

\subsection{Historical resource accounting}
We also recomputed the earlier Flash round-one usage totals from its 22 sessions per condition. Fresh input totals are 599,882 and 678,101; output totals are 116,506 and 132,726; cache totals are 5,890,176 and 5,017,984. The recorded total-token sums are 6,606,564 and 5,828,811, respectively. These fields do not support an undifferentiated claim of a 2.2-fold token increase. Their accounting differs from the focal host's token field, so resource ratios are reported within configurations only.

\section{Judge panel provenance}\label{app:panel}
The panel protocol, analysis plan, blinded items, unblinding key, and all verdicts are under \path{paper/audit/llm-judge-panel-v1}; \path{prepare-llm-judge-panel.mjs} and \path{analyze-llm-judge-panel.mjs} in \path{paper/scripts} regenerate the items and results and check them for drift. The protocol was written before any panel verdict existed but was not separately committed or externally registered beforehand. The Claude session that prepared the items had seen earlier results; it judged no item and passed no outcome information to the 16 judging sessions. The operator of the GPT-5.5 session had seen manuscript context before reading the blinding protocol and passed no outcomes to the isolated judge. Both judges' metadata files record these facts, including incidental directory-name listings in which no verdict or key file was opened. GPT-5.5's single sequential session may carry context between items, unlike the original per-report calls.

\section{Artifact map and reproduction scope}
Historical result files are stored under \path{benchmark/results}: \path{paired-effect-stats.json} generates the paired table, and \path{glm-pair-stability.json} records GLM stability. The focal directory contains the selection, schedule, execution-order records, original reports and judgments, aggregate scores, paired analysis, and the earlier targeted review. The later review stores frozen selection and complete criterion verdicts separately. The script \path{paper/scripts/summarize-submission-evidence.mjs} checks report hashes, recomputes review scores, deduplicates overlapping answers, and regenerates sensitivity and historical resource summaries without model calls. Its \texttt{--check} mode verifies the committed generated summary.

A frozen 23-task historical inventory and its source metadata are retained in \path{paper/generated}; they are not substituted for the 16-task selection. Exposure information remains in \path{paper/audit/task-exposure-ledger.csv}. Missing raw artifacts, unverified model identities, and pending author declarations are recorded as limitations rather than filled in by inference.
\end{document}

%% file: generated/benchmark-metadata.tex
\newcommand{\BenchmarkTaskCount}{23}

\newcommand{\BenchmarkStaticCount}{10}
\newcommand{\BenchmarkHandsOnCount}{13}

%% file: generated/contract-coverage-table.tex
\begin{table}[t]\centering\small
\begin{tabular}{p{0.37\textwidth}rrrr}
\toprule
Primary domain & Tasks & Criteria/arm & No skill & Skill\\\midrule
Version and release applicability & 6 & 12 & 8 & 12 \\
API, data and ownership contracts & 13 & 46 & 35 & 45 \\
Lifecycle, ordering and deployment & 9 & 26 & 24 & 25 \\
Safety and boundary handling & 5 & 10 & 10 & 10 \\
Failure attribution & 11 & 32 & 32 & 32 \\
Evidence, verification and provenance & 14 & 38 & 31 & 36 \\
\bottomrule\end{tabular}
\caption{Complete contract-stratified analysis of the focal 64 reports. The final two columns count original full-credit criterion decisions; they are not independently validated correctness counts. Criteria/arm includes both repeats. Multi-contract criteria receive one retrospective primary label; tasks can contribute to multiple domains.}
\label{tab:contracts}\end{table}

%% file: generated/contract-credit-figure.tex
\begin{figure}[t]
\centering
\begin{tikzpicture}
\begin{axis}[
  scale only axis, width=10.3cm, height=5.9cm,
  xmin=0, xmax=103, ymin=0.5, ymax=6.6,
  xtick={0,20,40,60,80,100}, ytick={6,5,4,3,2,1},
  yticklabels={{Version / release ($n=12$)},{API / data / ownership ($n=46$)},{Lifecycle / deployment ($n=26$)},{Safety / boundaries ($n=10$)},{Failure attribution ($n=32$)},{Evidence / verification ($n=38$)}},
  tick label style={font=\footnotesize},
  xlabel={Original full-credit decisions (\%)},
  xmajorgrids, grid style={gray!20}, axis lines*=left,
  legend style={at={(0.02,1.04)},anchor=south west,draw=none,font=\footnotesize,legend columns=2},
  clip=false]
\addplot[gray!60, thick, forget plot] coordinates {(66.66666666666667,6) (100,6)};
\addplot[gray!60, thick, forget plot] coordinates {(76.08695652173913,5) (97.82608695652173,5)};
\addplot[gray!60, thick, forget plot] coordinates {(92.3076923076923,4) (96.15384615384616,4)};
\addplot[gray!60, thick, forget plot] coordinates {(100,3) (100,3)};
\addplot[gray!60, thick, forget plot] coordinates {(100,2) (100,2)};
\addplot[gray!60, thick, forget plot] coordinates {(81.57894736842105,1) (94.73684210526316,1)};
\addplot[only marks,mark=o,mark size=3.6pt,very thick,color=oiblue] coordinates {(66.666667,6) (76.086957,5) (92.307692,4) (100.000000,3) (100.000000,2) (81.578947,1)};
\addlegendentry{No skill (circle)}
\addplot[only marks,mark=triangle*,mark size=2.6pt,color=vermillion] coordinates {(100.000000,6) (97.826087,5) (96.153846,4) (100.000000,3) (100.000000,2) (94.736842,1)};
\addlegendentry{With skill (triangle)}
\end{axis}
\end{tikzpicture}
\caption{Original full-credit proportions rise in four contract domains and remain at 100\% in two. Points summarize all 328 criterion decisions across 64 reports; $n$ is the number of criterion decisions per arm, including both repeats. Circles and triangles overlap at 100\% for safety/boundaries and failure attribution. These are finite-cohort descriptions of the original judge, not correctness rates or independent samples; the reviewed S11 boundary defect remains despite full original boundary credit.}
\label{fig:contract_credit}
\end{figure}

%% file: generated/judge-sensitivity-figure.tex
\begin{figure}[t]
\centering
\begin{tikzpicture}
\begin{axis}[
  scale only axis,width=10.3cm,height=5.9cm,
  xmin=-1,xmax=21,ymin=0.5,ymax=5.6,
  xtick={0,5,10,15,20},ytick={5,4,3,2,1},
  yticklabels={{GLM (original)},{Claude Opus 5.5},{GPT-5.5},{All three: mean},{Claude + GPT: mean}},
  tick label style={font=\footnotesize},
  xlabel={Mean paired gain (reward points)},
  xmajorgrids,grid style={gray!20},axis lines*=left]
\addplot[black,dashed,forget plot] coordinates {(0,0.5) (0,5.5)};
\addplot[gray!60,densely dotted,forget plot] coordinates {(-1,2.5) (21,2.5)};
\addplot+[only marks,black,mark=square*,mark size=3pt,error bars/.cd,x dir=both,x explicit,error bar style={thick}] coordinates {(4.9219,5) += (5.937475,0) -= (4.6094,0)};
\addplot+[only marks,oiblue,mark=triangle*,mark size=3pt,error bars/.cd,x dir=both,x explicit,error bar style={thick}] coordinates {(10.625,4) += (8.515625,0) -= (7.1875,0)};
\addplot+[only marks,vermillion,mark=diamond*,mark size=3pt,error bars/.cd,x dir=both,x explicit,error bar style={thick}] coordinates {(6.0938,3) += (4.99995,0) -= (4.5313,0)};
\addplot+[only marks,gray!70!black,mark=o,mark size=3pt,error bars/.cd,x dir=both,x explicit,error bar style={thick}] coordinates {(7.2131,2) += (6.249399999999999,0) -= (5.23435,0)};
\addplot+[only marks,gray!70!black,mark=o,mark size=3pt,error bars/.cd,x dir=both,x explicit,error bar style={thick}] coordinates {(8.3588,1) += (6.601825,0) -= (5.7025500000000005,0)};
\end{axis}
\end{tikzpicture}
\caption{All three judges retain a positive mean gain, while its size depends on the judge. Filled markers show the original GLM score and the two complete LLM re-judgments of the same 64 reports; open markers below the dotted separator average judges within each report and are not additional independent judges. Whiskers are 95\% percentile intervals from 10,000 paired-task bootstrap samples over the same 16 tasks (seed 20260915). All rows share the original rubric and reports; these intervals describe task resampling, not uncertainty over judge populations.}
\label{fig:judge_sensitivity}
\end{figure}
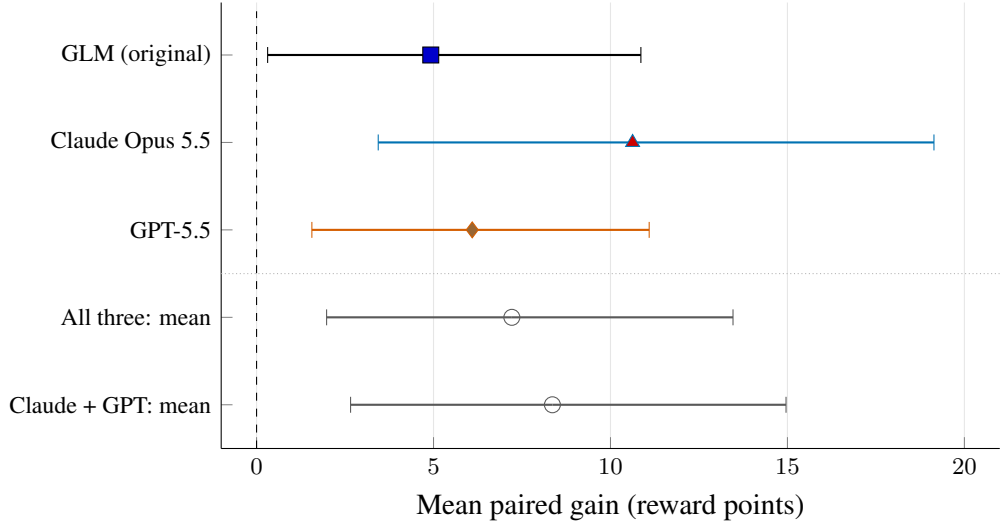

%% file: generated/llm-judge-panel-table.tex
\begin{table}[t]
\centering
\small
\setlength{\tabcolsep}{3pt}
\begin{tabular}{lccccc}
\toprule
Judge & no-skill $\rightarrow$ with-skill & $\Delta$ & 95\% CI & vs.\ original & vs.\ author \\
\midrule
  GLM-5.3-Flash (original) & 93.83 $\rightarrow$ 98.75 & +4.92 & [+0.31, +10.86] & -- & 53/56 \\
  Claude Opus 5.5 & 87.19 $\rightarrow$ 97.81 & +10.63 & [+3.44, +19.14] & 91.8\% / 0.64 & 51/56 \\
  GPT-5.5 & 93.28 $\rightarrow$ 99.38 & +6.09 & [+1.56, +11.09] & 95.7\% / 0.72 & 48/56 \\
  All-judge mean & 91.43 $\rightarrow$ 98.65 & +7.21 & [+1.98, +13.46] & -- & -- \\
  Cross-family mean & 90.23 $\rightarrow$ 98.59 & +8.36 & [+2.66, +14.96] & -- & -- \\
\bottomrule
\end{tabular}
\par\smallskip
{\footnotesize All 64 focal reports (328 criterion decisions) re-judged without supplied arm labels or prior scores under the unchanged report-judge-v2 prompt and rubric. ``vs.\ original'': exact credit-level agreement / linearly weighted $\kappa$ with the original GLM decisions. ``vs.\ author'': exact agreement with the 56 decisions of the non-blind plugin-author review (10 answers), which is a reference, not ground truth. Between panel judges: claude-opus-5-5 vs openai-gpt-5.5: 90.5\% exact, $\kappa_w$ = 0.57. All judges are LLMs; this is not human validation.\par}
\caption{Sensitivity of the focal paired endpoint to judge configuration. Estimand, bootstrap, and seed as in Section~\ref{sec:focal-method}; the original row is the primary recorded endpoint.}
\label{tab:judge_panel}
\end{table}

%% file: generated/focal-task-table.tex
\begin{longtable}{lrrrrr}
\caption{Focal task-level original scores. Each arm has two repeats; differences use arm means. These are static rubric scores, not repair pass rates.}\label{tab:focal_tasks}\\
\toprule
Task & No skill r1 & No skill r2 & Skill r1 & Skill r2 & Mean $\Delta$\\
\midrule\endfirsthead
\toprule Task & No skill r1 & No skill r2 & Skill r1 & Skill r2 & Mean $\Delta$\\\midrule\endhead
S1 & 55.00 & 60.00 & 100.00 & 100.00 & 42.50 \\
S2 & 100.00 & 80.00 & 100.00 & 100.00 & 10.00 \\
S5 & 87.50 & 100.00 & 100.00 & 100.00 & 6.25 \\
S6 & 87.50 & 87.50 & 100.00 & 100.00 & 12.50 \\
S9 & 100.00 & 100.00 & 100.00 & 100.00 & 0.00 \\
S11 & 100.00 & 90.00 & 90.00 & 90.00 & -5.00 \\
S12 & 100.00 & 100.00 & 100.00 & 100.00 & 0.00 \\
S13 & 100.00 & 100.00 & 100.00 & 100.00 & 0.00 \\
S14 & 100.00 & 100.00 & 100.00 & 100.00 & 0.00 \\
S15 & 100.00 & 100.00 & 100.00 & 100.00 & 0.00 \\
S17 & 80.00 & 80.00 & 90.00 & 100.00 & 15.00 \\
S18 & 100.00 & 100.00 & 100.00 & 90.00 & -5.00 \\
S19 & 100.00 & 100.00 & 100.00 & 100.00 & 0.00 \\
S20 & 97.50 & 97.50 & 100.00 & 100.00 & 2.50 \\
S21 & 100.00 & 100.00 & 100.00 & 100.00 & 0.00 \\
S22 & 100.00 & 100.00 & 100.00 & 100.00 & 0.00 \\
\bottomrule
\end{longtable}

%% file: generated/paired-effect-table.tex

\begin{table*}[t]
\centering
\small
\setlength{\tabcolsep}{3pt}
\begin{tabular}{>{\raggedright\arraybackslash}p{0.17\textwidth}>{\raggedright\arraybackslash}p{0.13\textwidth}>{\raggedright\arraybackslash}p{0.17\textwidth}>{\raggedright\arraybackslash}p{0.09\textwidth}>{\raggedright\arraybackslash}p{0.13\textwidth}>{\raggedright\arraybackslash}p{0.08\textwidth}>{\raggedright\arraybackslash}p{0.11\textwidth}}
\toprule
Model ($n$ tasks) & Tasks & no-skill $\rightarrow$ with-skill & Mean paired $\Delta$ & 95\% CI & Signed-rank $p$ & Protocol \\
\midrule
  qwen3.8-27b ($n = 56$) & full 56-task pool & 44.67 $\rightarrow$ 41.60 & $-$3.07 & [$-$8.13, +1.99] & 0.1771 & 3 scored, mean \\
  deepseek-v4-flash ($n = 23$) & 23-task snapshot & 69.96 $\rightarrow$ 80.63 & +10.67 & [+0.48, +22.11] & 0.1007 & 3-run median \\
  gpt-5.6-terra ($n = 21$) & 22-task pool, 21 scored & 71.10 $\rightarrow$ 79.76 & +8.67 & [$-$4.29, +21.62] & 0.2359 & single-shot \\
  glm-5.3-flash ($n = 22$) & S1--S22 static & 77.77 $\rightarrow$ 87.05 & +9.27 & [+3.95, +16.09] & 0.0056 & 3-round median \\
  glm-5.2 ($n = 22$) & S1--S22 static & 93.32 $\rightarrow$ 96.36 & +3.05 & [0.00, +7.18] & 0.1378 & 3-round median \\
\bottomrule
\end{tabular}
\par\smallskip
{\footnotesize Judge identity: in both GLM groups the solver and the grader belong to the same model family: glm-5.3-flash graded its own runs, and 19 of the 22 glm-5.2 tasks were judged by glm-5.3-flash subagents (the remaining three by official keyword judges), so correlated grader bias cannot be excluded (Section~\ref{sec:related}). glm-5.2's CI lower bound touching zero alongside $p = 0.1378$ is a percentile-bootstrap boundary artifact: with 16 of 22 task deltas tied at zero, the lower endpoint of the resampling distribution sits at zero, so the interval brushing zero does not contradict the nonsignificant test.\par}
\caption{Task-level paired effect of the plugin-upgrade skill across five historical configurations, ordered by each group's no-skill baseline (an outcome measure, not an independent capability metric; Section~\ref{sec:paired}). Each cell compares the two conditions on a 0--100 scale: per-task medians of three rounds (glm groups) or three runs (deepseek-v4-flash), per-task means of three scored attempts (qwen3.8-27b, reward means rescaled by 100), or single-shot rewards (gpt-5.6-terra; H8 excluded after verifier timeouts on both arms). Mean paired $\Delta$ is the mean of per-task with-skill-minus-no-skill deltas; 95\% CIs are percentile intervals from 10,000 task-level paired bootstrap replicates (mulberry32, seed 20260907); $p$ is the two-sided Wilcoxon signed-rank test (zero deltas excluded, tie-corrected normal approximation with continuity correction). Task pools and protocols differ across rows, so cross-row comparisons are descriptive.}
\label{tab:paired_effect}
\end{table*}